\documentstyle[preprint,aps]{revtex}

\begin{document}

\preprint{UCDPHY/ 96-33, USACH/96/09, USM-TH-68}

\title{ANOMALY AND CONDENSATE IN THE LIGHT-CONE SCHWINGER MODEL}
\author{J. GAMBOA$^{1,2}$\thanks{E-mail: jgamboa@lauca.usach.cl },  
I. SCHMIDT$^3$\thanks{E-mail: ischmidt@newton.fis.utfsm.cl} and 
L. VERGARA$^2$\thanks{E-mail: lvergara@lauca.usach.cl}}
\address{$^1$ Department of Physics, University of California, 
Davis, CA 95616, USA \\ 
$^2$ Departamento de F\'{\i}sica, Universidad de Santiago 
de Chile, 
Casilla 307, Santiago 2, Chile 
\\ $^3$ Departamento de F\'{\i}sica, Universidad T\'ecnica 
Federico Santa Maria, Casilla 110 - V \\ Valpara\'{\i}so,  Chile}
%\date{\today}
\maketitle

\begin{abstract} 
The axial anomaly and fermion condensate in
the light cone Schwinger model are studied, following path integral 
methods. This formalism allows for a simple and direct calculation of 
these and other vacuum dependent phenomena.

\end{abstract}

{\centerline {\bf 1. Introduction}}

\medskip
Light-cone quantization has been a very succesful formalism for 
calculating a number of physical quantities, mainly within 
perturbation theory \cite{Brodsky}.
In fact, in bound state calculations it has clear 
advantages over the usual equal-time formalism. In light-cone 
quantization all constituents have positive longitudinal momentum and 
energy $p^\pm ={1\over\sqrt{2}}(p_0\pm p_3)$, which means that the 
ground state of the system can have no constituents and it is then 
trivial. Thus a Fock-state expansion, in which one builds for example 
a constituent quark model in QCD which includes gluon and quark-antiquark 
emission, makes sense. Interactions which connect the perturbative 
vacuum which has $p^+ > 0$ to states with particles (each of these 
particles with positive $p^+$), are just not present. On the other    
hand, in equal-time quantization the vacuum is an infinite sea of 
constituents, and one cannot identify these excitations with a single 
quark-antiquark pair or a single triplet of quarks separate from the 
sea.

The quantization of gauge theories in the light cone has
been extensively studied in the last years in connection with
the possible understanding of non-perturbative phenomena
\cite{Pauli}, \cite{Perry}. The question immediately arises, however, 
of how it is possible that a trivial vacuum can give rise to 
phenomena, such as chiral symmetry breaking, condensate formation, and 
others, that are generally associated to a non-trivial vacuum 
structure. This issue has been studied recently, and one usually      
finds that it is neccesary to introduce an infrared regulator, but it 
is difficult to find a regulator that does not automatically remove 
the vacuum structure \cite{Robertson}.

In this paper we propose a simple direct method for defining a quantum 
theory in the light-cone, through a light-cone functional 
integral \cite{Gamboa}.
We will show that this formalism allows for a direct 
calculation of several non-perturbative vacuum associated phenomena.
We consider the massless Schwinger model which is a simple model
where many non-perturbatives properties, such as the axial
anomaly and the appearence of a fermion condensate, can be
studied\cite{Ji}.

The Schwinger model is described by the following action
\begin{equation} 
S_{SM} = \int d^2x [{\bar \psi} i{\gamma^\mu D_\mu}\psi - {1\over 4} 
F_{\mu \nu}F^{\mu \nu}], \label{sch} 
\end{equation}
with the covariant derivative $D_\mu$ defined as 
\begin{equation} 
D_\mu = \partial_\mu - ieA_\mu. \nonumber
\end{equation} 
\noindent The Dirac matrices satisfy
\begin{equation} 
\{ \gamma^\mu , \gamma^\nu \} = 2 \eta^{\mu \nu}, \nonumber
\end{equation} 
with $\eta^{\mu \nu} = diag (1, -1)$ and the $\gamma$-matrices
are represented by 
$\gamma^0 = \sigma_1, \gamma^1 = -i \sigma_2$ and 
$\gamma^5 = \gamma^0 \gamma^1$. 

In terms of light cone coordinates 
\begin{equation} 
x^\pm = {1\over \sqrt{2}} (x^0 \pm x^1), 
\end{equation} 
the $\gamma$-matrices and the covariant derivative become 
\begin{equation} 
\gamma^\pm = {1\over \sqrt{2}} (\gamma^0 \pm \gamma^1),
\,\,\,\,\,\,\,\, D_\pm =  {1\over \sqrt{2}} (D_0 \pm D_1). 
\end{equation}
\noindent Using this notation, the Lagrangian becomes 
\begin{equation} 
{\cal L} = {\bar \psi} (i \gamma_-D_+  + i\gamma_+D_-) \psi - 
{1\over 2} 
F^2_{+-}, \label{cono} 
\end{equation} 
\noindent and the formal partition function for this system is 
\begin{equation} 
{\cal Z }= {\displaystyle \int} {\cal D} A_+{\cal D} A_-
{\cal D}{\bar \psi}
{\cal D} \psi \,\,\, {\displaystyle e}^{iS_{SM} ( {\bar \psi}, \psi,
A_-, A_+)}. \label{formal}
\end{equation} 

Here we are using a Lagrangian quantization in order to right down the 
functional integral, which can be shown explicitely to be equivalent 
to a Hamiltonian functional integral quantization \cite{Gamboa}, 
\cite{Gitman}.

\medskip
{\centerline {\bf 2. The Anomaly}}
\medskip

In order to compute the axial anomaly we will introduce into
the partition function an auxiliary field $A^5_\pm$, defined
through the replacement 
\begin{equation} 
D_\pm \rightarrow {\cal D}_\pm = D_\pm - i A^5_\pm \gamma^5.
\label{cov} 
\end{equation}

Thus, the axial current can be
calculated by deriving the fermionic partition function with
respect to $A^5_\pm$ and then setting this field to zero at
the end of the calculation. 
However, before proceeding this way the fermionic partition
must be properly normalized such that the zero modes are
cancelled out. Therefore, instead of (\ref{formal}) our
starting point will be the partition function  
 
\begin{equation} 
Z = \frac{ {\displaystyle \int} {\cal D}{\bar \psi} {\cal D} \psi
\,\,\, {\displaystyle e}^{iS ( {\bar \psi}, \psi, A_-, A_+,
A^5_-, A^5_+) }}
{{\displaystyle \int} {\cal D}{\bar \psi} {\cal D} \psi
\,\,\, {\displaystyle e}^{iS ( {\bar \psi}, \psi)}}, \label{fer1}
\end{equation} 
which can be directly evaluated to give the ratio of two
determinants. They are in turn expanded and the constant part 
left out. Therefore, after using the explicit form of
$D_\pm$ and ${\cal D}_\pm$, one finds

\begin{eqnarray}
Z &=& Tr \{(e\gamma_- A_+ + \gamma_-A^5_+ \gamma^5 
+ e\gamma_+ A_- + \gamma_+ A^5_-\gamma^5)\cdot S_F\}
\nonumber \\ && -{1\over 2} Tr \{(e\gamma_- A_+ + \gamma_-
A^5_+ \gamma^5 + e\gamma_+ A_- + \gamma_+ A^5_-\gamma^5)\cdot
S_F  \nonumber \\&& \times (e\gamma_- A_+ + \gamma_- A^5_+
\gamma^5 + e\gamma_+ A_- + \gamma_+ A^5_-\gamma^5)\cdot S_F\},
\label{parti}
\end{eqnarray}
where $S^{-1}_F=i(\gamma_-\partial_+ + \gamma_+ \partial_-)$ and the
$Tr$ symbol means an integration over space
coordinates and a trace over gamma matrices. Higher order
terms should not contribute to the anomaly according to 
the Adler - Bardeen conjecture\cite{Adler}.
 
The divergence of the axial current is given by
 
\begin{equation}
<\partial \cdot j^5>= \partial_+ \frac{\delta Z}{\delta A^5_+}
\bigg\vert_{A^5_\pm =0} + \partial_- \frac{\delta Z}{\delta A^5_-}
\bigg\vert_{A^5_\pm =0},
\end{equation}
where $\partial \cdot j^5= \partial_+j^5_- +\partial_-j^5_+$.

One readily observes that by translation invariance the first term
in eq. (\ref{parti}) does not contribute to the anomaly.
Therefore, by going to Fourier space one finds that

\begin{eqnarray} 
<ik \cdot j^5> &=& - e \int \frac{dp_+dp_-}{(2 \pi)^2}
tr\{\gamma^5 \frac{1}{\gamma_+
p_- + \gamma_- p_+}[\gamma_- A_+ + \gamma_+ A_-] \frac{1}{\gamma_+
(p_- - k_-) + \gamma_- (p_+ - k_+)}
\nonumber \\ 
&&\times [\gamma_- k_+ + \gamma_+ k_-]\}, \nonumber \\ 
&=& 
e \int \frac{dp_+dp_-}{(2 \pi)^2} \biggl\{ tr \biggl[ \gamma^5 
\frac{1}{\gamma_+
p_- + \gamma_- p_+}[\gamma_- A_+ + \gamma_+ A_-]\biggr] -
\nonumber \\  
&& tr \biggl[ \gamma^5 
\frac{1}{\gamma_+
(p_- - k_-) + \gamma_- (p_+ - k_+) }[\gamma_- A_+ + 
\gamma_+ A_-]\biggr]\biggr\}, \label{int}
\end{eqnarray}
where the identity
\begin{equation} 
\gamma_- k_+ + \gamma_+ k_- = - \gamma_+ (p_- - k_-) - 
\gamma_- (p_+ - k_+) + \gamma_- p_+ + \gamma_+ p_- 
\nonumber
\end{equation} 
has been used.
The integrals that appears in (\ref{int}) are linearly
divergent and as is well known one cannot make a shift in the
momenta. The integrals can be performed by expanding the
second term around $k=0$, {\it i.e.} 
\begin{eqnarray} 
\frac{1}{ {\gamma_+ (p_- - k_-) + 
\gamma_- (p_+ - k_+)}} 
&=& 
\frac{1}{{\gamma_+p_- + \gamma_- p_+} } 
\nonumber \\&&+ \frac{1}{{\gamma_+p_- + \gamma_- p_+} } 
(\gamma_- k_+ + \gamma_+ k_-) 
\frac{1}{{\gamma_+p_- + \gamma_- p_+} }, 
\end{eqnarray} 
where we have not written the  higher order terms because they 
do not contribute to the integrals (\ref{int}). 
In addition those integrals are 
infrared divergent and must be regularized by adding a mass term in 
the propagator. 

After this regularization one finds that
\begin{eqnarray} 
<ik \cdot j^5> &=& - e \int \frac{dp_+dp_-}{(2 \pi)^2} tr 
\biggl\{\gamma^5 \frac{1}{\gamma_+
p_- + \gamma_- p_+ -m}(\gamma_- k_+ + \gamma_+ k_-) \nonumber \\
&& \times \frac{1}{\gamma_+p_- 
+ \gamma_- p_+} ({\gamma_-A_+ + \gamma_+A_-}) \biggr\} \nonumber 
\\ 
&=&  - e \int \frac{dp_+dp_-}{(2 \pi)^2} 
\frac{1}{({{2p_+p_- - m^2}})^2} \biggl[ tr \bigg\{ \gamma^5({\gamma_+ p_- 
+  \gamma_- p_+})({\gamma_+ k_- 
+  \gamma_- k_+}) 
\nonumber 
\\ && \times({\gamma_+ p_- +  \gamma_- p_+}) ({\gamma_+ A_- 
+  \gamma_- A_+})\biggr\} \nonumber \\&&
+ m^2 tr \biggl\{ \gamma^5 ({\gamma_+ k_- + \gamma_- k_+})
({\gamma_+ A_- + \gamma_- A_+}) \biggr\} \biggr], \label{anoma1} 
\end{eqnarray} 
and using standard identities for the $\gamma$-matrices 
(\ref{anoma1}) reads
\begin{equation} 
<ik \cdot j^5> = - e \int \frac{dp_+dp_-}{(2 \pi)^2}  
\frac{1}{{(2p_+ p_- - m^2)}^2}
\biggl[ 4 {(p^2_- k_+ A_+ - p^2_+ k_- A_-)} + 2 m^2
{(k_- A_+ - k_+ A_-)} \biggr]. \label{anoma2} 
\end{equation}  

The first term on the RHS vanishes by symmetry considerations. In fact, 
by writing 
\begin{equation} 
\int \frac{dp_+dp_-}{(2 \pi)^2}  
\frac{{(p^2_- k_+ A_+ - p^2_+ k_- A_-)} }{{(2p_+ p_- - m^2)}^2} 
= {(k_+ A_+ - k_- A_-)}I , \label{sim1}
\end{equation} 
with 
\begin{equation} 
I= \int \frac{dp_+dp_-}{(2 \pi)^2} \frac{p^2_\pm}{{(2p_+
p_- -m^2)}^2}, \label{def1} 
\end{equation} 
from Eqs. (\ref{sim1}) and (\ref{def1}) one notes that by interchanging 
$(k_+, A_+) \leftrightarrow (k_-, A_-)$ one gets 
\begin{equation} 
(k_+ A_+  -  k_- A_- ) I = -(k_+ A_+ - k_- A_- ) I = 0.
\nonumber 
\end{equation} 

The integral on the second term on the RHS in (\ref{anoma2}) can be
directly evaluated to give 
\begin{equation} 
 \int \frac{dp_+dp_-}{(2 \pi)^2}  
\frac{1 }{{(2p_+ p_- - m^2)}^2} = \frac{i}{4\pi m^2},  \label{anom4}
\end{equation}  
and therefore the expectation value of the divergence of the axial
current in momentum space is 
\begin{equation} 
< k\cdot j^5> = \frac{e}{\pi} {(k_+ A_- - k_- A_+)}, \nonumber
\end{equation} 
or in coordinate space 
\begin{equation} 
< \partial \cdot j^5 > = -\frac{e}{\pi} F_{+-}, \label{anom5} 
\end{equation} 
which is the axial anomaly. 

\medskip
{\centerline {\bf 2. The Condensate and Mass Gap}}
\medskip

The same arguments presented above remain valid when one computes
the fermion condensate. In fact, the central point is to find an
argument that permits us to compute $<{\bar \psi} \psi>$
independently of $x$, but as is well known, this quantity can be
computed
in quantum field theory by looking at an equivalence between
$<{\bar \psi} \psi>$ and $G(x,0)$ defined as  
\begin{equation} 
G(x,0) = <0|{\bar \psi}(x) \psi (x){\bar \psi}(0) \psi (0)|0 >,
\label{cond1} 
\end{equation} 
where $|0>$ is the perturbative vacuum. Then, following standard
arguments, one evaluates $G(x,0)$ when $x$ goes to infinity
\footnote{Here one should note that such a limit is well defined
in euclidean space. There $x_+ \rightarrow z=ix_0 + x_1$ and
$x_- \rightarrow {\bar z}=-ix_0 + x_1$.}
and finds that
\begin{equation} 
\lim_{x \rightarrow \infty} G(x,0) = \frac{1}{2} {\vert <{\bar
\psi} \psi> \vert}^2. \label{cond2}
\end{equation}
 
This last formula is heavily dependent of a non-trivial vacuum
and the problem is, how can one correlate it with the light cone
vacuum. A possible answer to this question has been given in
references \cite{Robertson,Heinzl}. Essentially these authors find 
that the above formulae remain valid in the light cone and here we 
suppose that this is true. 

The path integral representation of $G(x,0)$ for the Schwinger model is    
\begin{equation} 
G(x,0)  = \frac{{\displaystyle \int} {\cal D} A_\mu {\cal D}
{\bar \psi} {\cal D} \psi~  
{\bar \psi}(x) ~ \psi (x)~ {\bar \psi}(0) ~ \psi (0) \,\,\,
{\displaystyle e}^{i S_{SM}}}
{{\displaystyle \int} {\cal D} A_\mu {\cal D} {\bar \psi}
{\cal D} \psi  \,\,\,
{\displaystyle e}^{i S_{SM}}}, \label{G1}
\end{equation} 
where $S_{SM}$ is defined in (\ref{sch}). 

In the light cone, the gauge field can be decomposed as 
\begin{equation} 
A_\pm =  \mp \partial_\pm S + \partial_\pm \varphi, 
\label{a}
\end{equation}
where $S$ and $\varphi$ are pseudoscalar and scalar fields
respectively. Then, if one makes a chiral transformation  
\begin{equation} 
\psi = e^{-i\gamma^5 S} \psi^{'}, \,\,\,\,\,\,\,\,\,\,\,\,\, 
{\bar \psi} = {\bar \psi^{'}} e^{-i\gamma^5 S}, \label{ch1}
\end{equation}
the functional measure is modified, {\it i.e.}
\begin{equation} 
{\cal D} {\bar \psi} {\cal D} \psi \rightarrow {\cal D}
{\bar \psi} {\cal D} \psi \,\,\,
{\displaystyle e}^{\frac{i}{\pi} \int dx_+ dx_- S\partial_+
\partial_- S}. \label{mea1} 
\end{equation} 
If in addition one performs a gauge transformation
\begin{equation}  
\psi = e^{i\varphi} \psi^{'}, \,\,\,\,\,\,\,\,\,\,\,\,\, 
{\bar \psi} = {\bar \psi^{'}} e^{-i\varphi}, \label{g1} 
\end{equation} 
one easily sees that the fermionic fields decouple.Therefore,
the effective action that appears from these operations is
given by
\begin{equation}
S_{eff}=\int dx_+ dx_-\biggl( \frac{2}{e^2} {(\partial_+
\partial_-S)}^2 + \frac{1}{\pi} S(\partial_+
\partial_-S) + {\bar \psi} i \gamma
\cdot \partial \psi \biggr).\label{act}
\end{equation} 
In addition, the product of the four fermions   
${\bar \psi}(x) \psi (x){\bar \psi}(0) \psi (0)$ is changed by
the chiral transformation to
\begin{equation}
{\bar \psi}(x)\;e^{-2i
\gamma^5 S(x)} \psi (x)~{\bar \psi}(0)\;e^{-2i
\gamma^5 S(0)} \psi (0),
\end{equation}
and as a consequence, $G(x,0)$ becomes
\begin{equation} 
G(x,0)  = \frac{{\displaystyle \int} {\cal D} S {\cal D}
{\bar \psi} {\cal D} \psi  
~{\bar \psi}(x)~e^{-2i
\gamma^5 S(x)} ~ \psi (x)~{\bar \psi}(0)~{\displaystyle e}^{-2i
\gamma^5 S(0)}~ \psi (0) \,\,\, 
e^{ iS_{eff}}}{{\displaystyle \int} {\cal D} S {\cal D} {\bar \psi}
{\cal D} \psi  \,\,\, 
{\displaystyle e}^{iS_{eff}}}, \label{G3}
\end{equation} 
where a trace over the fermionic indices is implied. 
Here one should note that the longitudinal modes ($\varphi$) are exactly 
cancelled in the light cone without having used a gauge fixing
condition as is mandatory in the standard approach.

The integration over the fermions can be straightforwardly done and 
we find that

\begin{equation}
G(x,0)= - \frac{1}{4\pi^2 (x_+x_-)} \cdot 
\frac{{\displaystyle \int} {\cal D} S  \,\,\, 
{\displaystyle e}^{ iS_{eff}} 
\cos {(2[S(x) - S(0)])}}{{\displaystyle \int} {\cal D} S  \,\,\, 
{\displaystyle e}^{iS_{eff}}}.
\end{equation}  
These integrals are Gaussian and can be best evaluated in Euclidean
Fourier space,

\begin{equation}
G(|z|,0)= \frac{1}{4\pi^2 {|z|}^2} e^{ P(|z|)},
 \label{G4}
\end{equation}
where
\begin{eqnarray}
P(|z|)&=&\int \frac{dp d{\bar p}}{(2\pi)^2} \frac{{\big\vert
e^{ip\cdot z} -1 \big\vert}^2 }
{{|p|}^2 (2 {|p|}^2 + m^2)}\nonumber \\
&=&2 \{\gamma_E + \frac{1}{2}\ln(m\sqrt{2 {|z|}^2})
+ K_0(m\sqrt{2{|z|}^2})\}
\end{eqnarray}
with $m^2=e^2/\pi$ the mass of the photon (mass gap). $\gamma_E$ 
is the Euler constant and $K_0$ the associated Bessel function.

Therefore, in the limit $|z| \rightarrow \infty$ the fermion
condensate is
\begin{equation}
{\vert <{\bar \psi} \psi> \vert}^2 = \frac{m^2}{4\pi^2}
{\displaystyle e}^{2\gamma_E}.
\end{equation}
The same result can also be obtained by summing fermion-
antifermion diagrams as is discussed in \cite{Kovacs} in the
standard approach.

In conclusion we have discussed a functional integral approach
to light cone gauge theories. This approach has
technical advantages with respect to the standard light cone
approach because it allows a direct calculation of the anomaly
and the fermion condensate, avoiding the technicalities
present when one uses canonical light-cone quantization.

\medskip
\centerline{\bf ACKNOWLEDGEMENTS}
\medskip

This work was supported in part by Fondecyt (Chile) contracts 1950278, 
1960229, 1960536 and DICYT. 
L.V. was supported in part by Dicyt (USACH) 049631VC.
One of us (J. G.) is a recipient of a John S. Guggenheim Fellowship and 
he thanks also to S. Carlip the hospitality in UCD.

\end{document}